\documentclass[usenatbib]{mnras}
\usepackage{graphicx}

\usepackage{multirow}
\usepackage{bigdelim}

\title[H$_2$ flows in Mon R1 association]{Near-infrared detection of H$_2$ flows in the core of Mon R1 association\thanks{{\it Herschel} is an ESA space observatory with science instruments provided by European-led Principal Investigator consortia and with important participation from NASA.}}
\author[T. Yu. Magakian, A. M. Tatarnikov, T. A. Movsessian, and H.R Andreasyan]{T. Yu. Magakian$^{1}$\thanks{E-mail:tigmag@sci.am}, A. M. Tatarnikov$^{2}$, T. A. Movsessian$^{1}$ and H.R Andreasyan$^{1}$ 
\\
$^{1}$Byurakan Astrophysical Observatory,  Aragatsotn reg.,Armenia\\
$^{2}$M. V. Lomonosov Moscow State University, Sternberg Astronomical Institute,
119991, Universitetskij pr. 13, Moscow, Russia }
\usepackage{amssymb}

\begin{document}

\date{}

\pagerange{\pageref{firstpage}--\pageref{lastpage}} \pubyear{2002}

  \maketitle

\label{firstpage}

\begin{abstract}

We report the discovery of 4 new H$_2$ jets in Mon R1 star-forming region on the images obtained with the 2.5-m telescope
of the Caucasian Mountain Observatory of SAI MSU through the filter, centered on the H$_2$ 1-0 S(1) emission line. This discovery confirms the nature of these flows, which existence was previously suspected using archival Spitzer GLIMPSE360 and WISE survey images. Also two infrared reflection nebulae were revealed. On the Herschel PACS survey images we found a small group of far-infrared sources, mostly unknown ones. Among them are the possible exciting objects of these outflows. Spectral energy distributions of new sources show their extremely red colour and the bolometric luminosities reaching 3 L$_{\sun}$ and  even 10L$_{\sun}$. They should be PMS objects at the very early evolutionary stages.
\end{abstract}


\section{Introduction}

Mon R1 association \citep{Racine} is a group of stars, illuminating bright reflection nebulae of various sizes, located to the north-west from the well-known Mon~OB1 association. In the updated list  about 30  optically visible young stellar objects (YSOs) were included as the Mon~R1 members \citep{Herbst}.  Nearly all more recent observational studies in the Mon~R1 field were related to  the small cluster of probable YSOs around VY~Mon \citep[and references therein]{WL,gutermuth}, the luminous  HAeBe star, illuminating the IC~446 reflection nebula.

\begin{figure}
        \includegraphics[width=250pt]{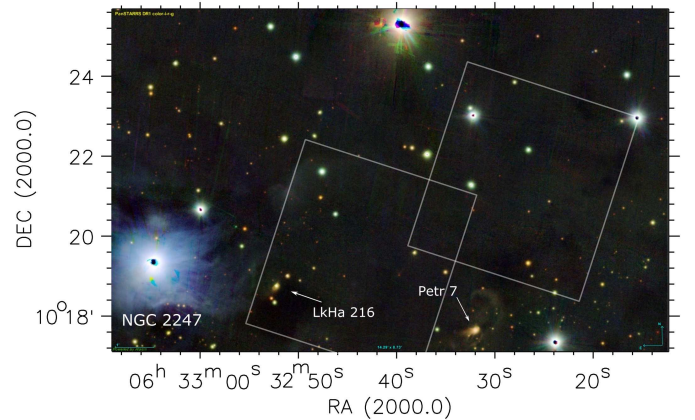}
        \caption{General view of the studied field in the PanSTARRS color image. Bright nebulosities are marked. White rectangles show two observed positions. }
        \label{field}
\end{figure}

In our previous paper \citep{MMD} we described the results of a search of Herbig-Haro (HH) objects in the field, centered on the three brightest nebulae NGC~2245, NGC~2247
and IC~446 in the Mon~R1 association. Many new HH flows and their probable sources were found, significantly increasing the number of YSOs in Mon~R1. In the course of this work we noted several elongated nebulous knots, arranged in chains and visible mainly in 4.5 $\mu$m bands both of  Spitzer and WISE surveys, but not detected in 3.6 $\mu$m. It was natural to assume that they can represent molecular H$_2$ flows. To check this assumption we organized their observations with 2.5 m telescope of Sternberg astronomical institute.

\section[]{Observations}

The narrow-band filter observations of Mon R1 association region were
carried out with the 2.5-m telescope
of the Caucasian Mountain Observatory of SAI MSU \citep{Shatsky} in 2020.
We obtained images of two fields in the $v =$ 1-0\ $S(1)$\ H$_2$ band ($\lambda_c$ = 2.132 $\mu$m,
bandwidth = 0.046 $\mu$m) and
``Kcont'' band
($\lambda_c$ = 2.273 $\mu$m, bandwidth = 0.039 $\mu$m) with the ASTRONIRCAM
infrared
camera-spectrograph \citep{Nadjip}. The field of view and image scale
were about 4.5\arcmin\ and 0.27\arcsec/pixel respectively.
The final images are the result of sum a number of individual images with a total exposure time of
about 2000 s.

\section{Results}

In Fig.\ref{field} we show the general view of that part of Mon R1, where the suspected H$_2$ flows were found.  All of them are located in a very dense area, seen to the west from NGC 2247 nebula, near  IRAS~06297+1021 (W) and (E) sources \citep[see for details][]{MMD}.

\begin{figure*}
        \includegraphics[width=430pt]{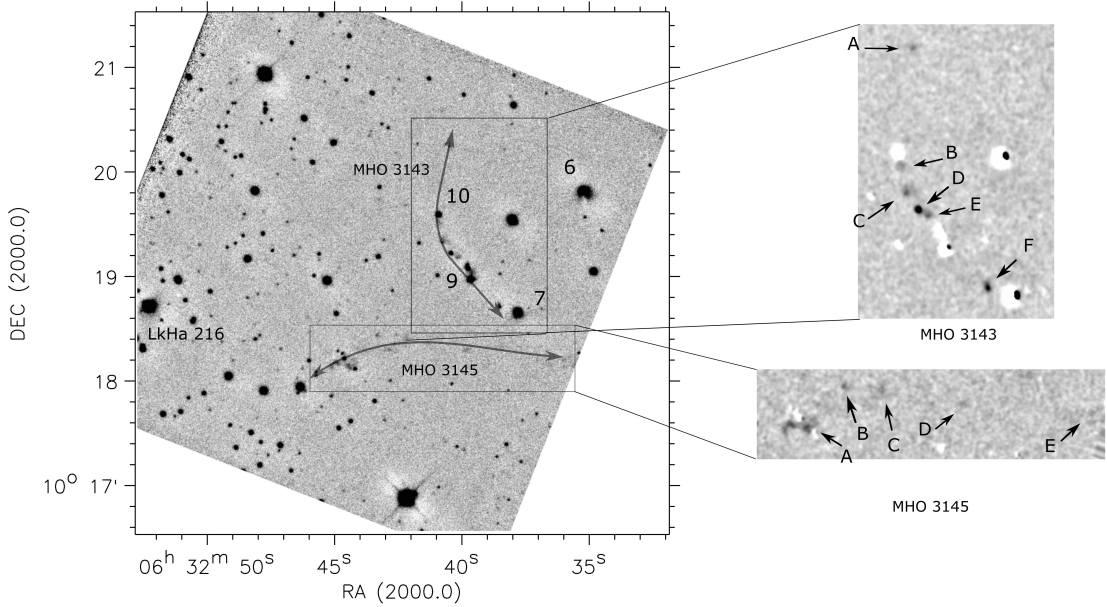}
        \caption{Flows MHO 3143 and MHO 3145. H$_2$ image is shown in left. The insets in the right show the details of the flows on H${_2} -$``Kcont'' images. The approximate extent and paths of the flows are marked by solid lines (left); the individual knots are marked by letters (right). Several stars, discussed in the Sec.4, are labeled by their numbers from Table 2 (left).}
        \label{Flows_E}
\end{figure*}

\begin{figure*}
        \includegraphics[width=430pt]{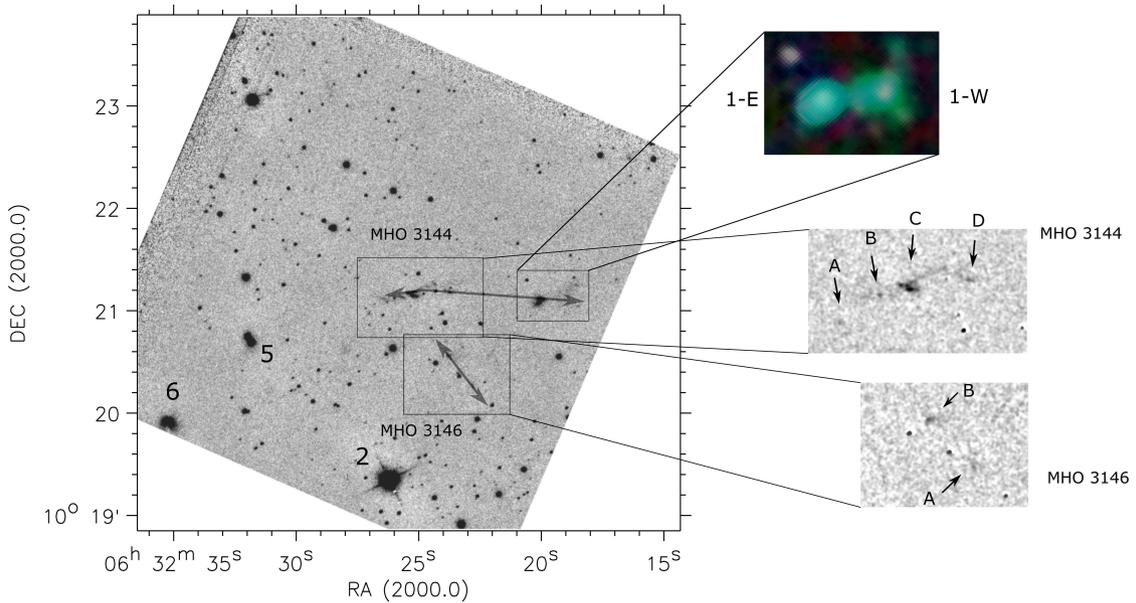}
        \caption{Flows MHO 3144 and MHO 3146. H$_2$ image is shown in left.  Two insets in the right show the details of the flows on H${_2} -$``Kcont'' images.  The upper right inset shows the magnified and contrast-enhanced image of the comet-like nebula, probably connected with MHO 3144, from GLIMPSE360 survey. Other marks and labels have the same meaning as in Fig. \ref{Flows_E}. }
        \label{Flows_W}
\end{figure*}

The field of probable H$_2$ flows was covered with two pair of images, obtained in 2.12 $\mu$m\ 1-0 S(1) H$_2$\ line and K-band continuum. Their approximate positions also are shown in Fig.\ref{field}. We present our findings in Figs. \ref{Flows_E} and \ref{Flows_W}. Reality of all three suspected molecular flows was confirmed; besides, the existence of one more flow, not well discernible on GLIMPSE360 images, was revealed. However, it was not so easy to point  their probable sources. Below we describe all four outflows separately.

\textbf{MHO 3143}. This is the very well defined and narrow outflow, elongated mainly in NE-SW direction, with some turns and wiggling. It consists of at least six compact knots of various brightness (see Fig.\ref{Flows_E}). The axis of the flow crosses several stars of various colors. One of them should be the probable source. Near the star 10, which seems the most promising candidate  (see the next section for discussion) the jet abruptly changes direction. In Table 1 we preliminarily denoted these parts as southern and northern lobes and measured PA for both of them.  

\textbf{MHO 3144}. This flow propagates in E-W direction.
In fact, it probably consists  of two extended patches, which include internal  structures (Fig.\ref{Flows_W}). The brightest part of the eastern patch (knot C) is split, making appearance of the disrupted bow-shock, heading to the east. Several small knots and nebular wisps can be seen eastward and westward from it, extending this patch to more than one arcuate minute. One cannot exclude that some of these knots can belong to MHO 3146 flow, which perhaps crosses this patch. The western patch of MHO 3144 consists of a comet-shape nebula with bright core. Its second, rather faint but distinctly visible nebulous cone with opposite orientation  can be seen westward from the core, making this nebula to appear as a bipolar object. In fact, however, each of the cones contains a bright in middle-IR range star; these stars and the nebula can be seen in the images of allWISE and unWISE surveys, as well as is the colour GLIMSE360 image, shown in Fig.\ref{Flows_W}.  The nebula probably is of reflection nature, being visible mainly in K-band continuum. One of its central stars naturally can be assumed as the source of the flow.  
 
\textbf{MHO 3145}.
 This flow is located to the south of MHO 3143 and, as MHO 3144, is directed in E-W. It is rather long and includes at least five distinct patches, but they are fainter than in the two flows described above. The eastern patch of the flow (knot A) is the brightest. It consists
of several small clumps (Fig.\ref{Flows_E}).
The westernmost component (knot E) is partly outside from our H$_2$ image. This flow has no obvious source. MHO 3143 and MHO 3145 can intersect each other.
 
\textbf{MHO 3146}.
This flow is faint and we noted it only on the H$_2$ images. Its axis is not well defined, but seems  nearly parallel to MHO 3143.
It consists of at least two diffuse patches and, perhaps, reaches MHO 3144.

   The question of the sources of these flows is discussed in the next section. 

In Table 1 we present the approximate coordinates of the centers of all four flows and their projected lengths in parsecs. We assumed 715 pc for the distance of Mon~R1 association \citep[see][for the further details]{MMD}. As one can see, visible sizes of the flows correspond to significant lengths -- up to nearly half of parsec.
     
\begin{table*}
 \centering
 \begin{minipage}{160mm}
  \caption{The coordinates and parameters of H$_{2}$ flows in the Mon~R1 field.}
  \begin{tabular}{lccccccc@{}}
  \hline
   No.     &  RA(2000)    & Decl.(2000) & Full projected & PA${^b}$ & \multicolumn{2}{c}{Flux (W/m${^2}$)$\times$ 10$^{-17}$}  & L$_{2.12}$$^{c}$ \\ 
   & h m s & \degr\ \arcmin\ \arcsec & length (pc)$^a$ & (degr) & obs & corr & $\times10^{-3}L_{\sun}$ \\

 \hline
MHO 3143 & 06 32 40.9 & +10 19 37 &  0.36 & 207, 351$^d$ & 54.72 &   103.15 & 16.44\\
MHO 3144 & 06 32 23.9 & +10 21 11 & 0.49 & 88 & 40.21 &  75.80 & 12.08\\
MHO 3145 & 06 32 39.8 & +10 18 19 & 0.43 & 89 & 29.68 & 55.95 & 8.92\\
MHO 3146 & 06 32 23.8 & +10 20 33 & 0.22 & 217 &  6.7 & 12.63 &  2.01 \\
 \hline

\end{tabular}

\begin{flushleft}
\textit{Notes}: \\
$^{(a)}$  Determined for 715 pc distance. \\
$^{(b)}$ Position angle of the outflows measured east of north. \\
$^{(c)}$ The
luminosity is calculated by summing the de-reddened for A$_{V}=6$  H$_{2}$ 1-0\ $S(1)$ line flux values for the individual components and
converting these fluxes into a luminosity assuming a distance of 715 pc. \\
$^{(d)}$ PA for southern and northern lobes \\

\end{flushleft}
\end{minipage}
\end{table*}

\section{Discussion}

To estimate the luminosity of the flows in H$_2$\ emission
we measured the flux for each of their individual knots, using several stars with 2MASS K$_{s}$ magnitudes in the field for calibration, and then summed them.
The relatively small area, which contains the newly found H$_2$\ flows, falls within the dark cloud LDN 1605, where extinction (A$_{V})$, according to  catalog of \citet{dobashi}, reaches 5 magnitudes, and according to the detailed maps of \citet{rf}, even 7.5 magnitudes in the certain points. To correct the fluxes for the reddening we assumed A$_{V}=6$ as the reasonable estimate. Then we computed the luminosity of each flow, using the  dereddened fluxes and 715 pc for the distance of Mon R1. These results also are presented in Table 1.  

As was mentioned above, several HH groups and HH flows were found in the six areas of Mon R1 and described in the our previous paper \citep{MMD}. It is worth mentioning that none of these areas  is located in the field, studied here, and, besides, none of these groups shows obvious traces of H$_2$\ emission at least in WISE or  GLIMPSE360 surveys. However, we can compare, for example, the results for MHO 3143-3146 with the data about Cyg OB7 association, where significant amount of HH and H$_2$\ flows were found \citep[][and references therein]{Khanzadyan}, and which is located in the similar distance (800 pc). As one can see, the luminosities of H$_2$\ flows in both associations are quite similar.

 In any case, the void, located eastward to NGC 2247 and visible in the optics (Fig.\ref{field}), becomes more or less filled in the 2MASS image. WISE and especially unWISE images show large amount of presumably  background stars nearly everywhere in the field of interest; the most extincted part of $2\arcmin \times 3\arcmin$ size seems to be located just to the north from MHO~3144. However, even on  the WISE and Spitzer GLIMPSE360 images the number of stars, which can be considered as the possible sources of the newly detected H$_2$ flows, is very low. One can suppose that at least part of possible source stars can have dense dust envelopes, thus being much more extincted.

\begin{figure*}
        \includegraphics[width=450pt]{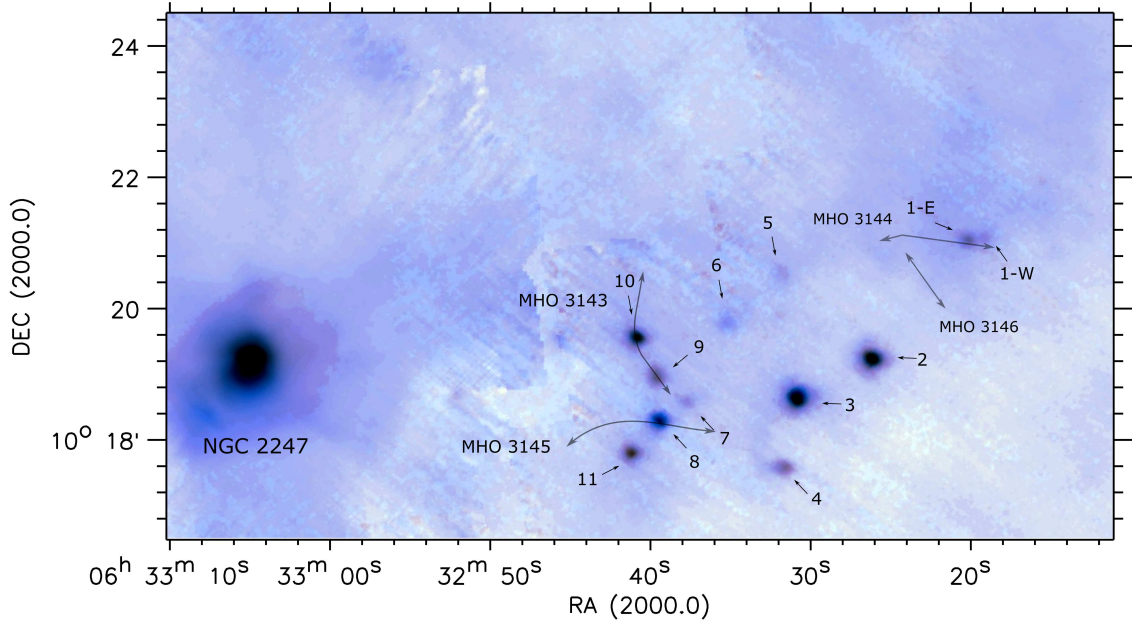}
        \caption{The studied field in Herschel PACS color image, which is built from images, obtained in  70 $\mu$m and 160 $\mu$m  wavelengths. Eleven detected sources are labeled by numbers. The positions, directions and approximate extent of MHO 3143-3146 flows also are shown.}
        \label{PACS}
\end{figure*}

To check this assumption we analysed the far-infrared PACS 70 and 160 micron images (100 micron survey did not cover our area), obtained with Herschel space observatory (Fig.\ref{PACS}). They immediately reveal the existence of a small cluster of several far-IR sources just in the center of the area of interest. Two brightest ones  coincide with luminous IRAS~06297+1021 (W) and IRAS~06297+1021 (E) objects \citep[see][]{MMD}. Beyond this field only the bright  HAeBe stars HD 259431, LkH$\alpha$ 215 and VY Mon with their reflection nebulae, as well as IRAS 06292+1029 source, also  described in our previous paper, can be seen. It is remarkable, that none of these brightest IR sources is associated with any of newly found flows.

\begin{table*}
\label{sources}
 \centering
 \begin{minipage}{160mm}
  \caption{The coordinates and identifications of far-IR sources.}
  \begin{tabular}{llll@{}}
  \hline
   No.     &  RA(2000)    & Decl.(2000) & Notes  \\
 \hline
 1-W & 06 32 19.05 & +10 21 09.8 & allWISE J063218.93+102109.8 \rdelim\}{2}{1cm}[Double (17\arcsec), nebulous, absent in 2MASS. MHO 3144 source?]\\
 1-E &  06 32 20.07 & +10 21 07.3 & allWISE J063220.03+102107.2  \\
 2 & 06 32 25.99 & +10 19 18.6 & IRAS~06297+1021 (W) (2MASS J06322611+1019184)\\
 3 & 06 32 30.85& +10 18 40.0 & IRAS~06297+1021 (E) (2MASS J06323082+1018396) \\
 4 & 06 32 31.60 & +10 17 35.2 & Petr 7 (2MASS J06323159+1017352)\\
 5a & 06 32 31.82 & +10 20 39.9 & 2MASS J06323180+1020390 \rdelim\}{2}{1cm}[Very red double (4.5\arcsec) star]   \\
 5b & 06 32 32.0 & +10 20 43.7 & 2MASS J06323194+1020426  \\
 6 & 06 32 35.17 & +10 19 51.1 & 2MASS J06323513+1019511. Bright in IR, nebulous star, not galaxy  \\
 7 & 06 32 37.73 & +10 18 40.5 & 2MASS J06323771+1018403. MHO 3143 source?\\
 8 & 06 32 39.47 & +10 18 21.2 & allWISE J063239.49+101819.4; absent in 2MASS. Extremely red knot. MHO 3145 source?\\
 9 & 06 32 39.63 & +10 18 59.8 & 2MASS 06323958+1018593, allWISE J063239.61+101901.1. Close double. MHO 3143 source?\\
 10 & 06 32 40.90 & +10 19 39.0 & 2MASS 06324087+1019373, allWISE J063240.86+101937.4.  MHO 3143 probable source?\\
 11 & 06 32 41.34 & +10 17 49.0 & allWISE J063241.24+101750.6; absent in 2MASS. Extremely red. Nebulous?   \\ 
 \hline

\end{tabular}
\end{minipage}
\end{table*}

\begin{figure*}
        \includegraphics[width=400pt]{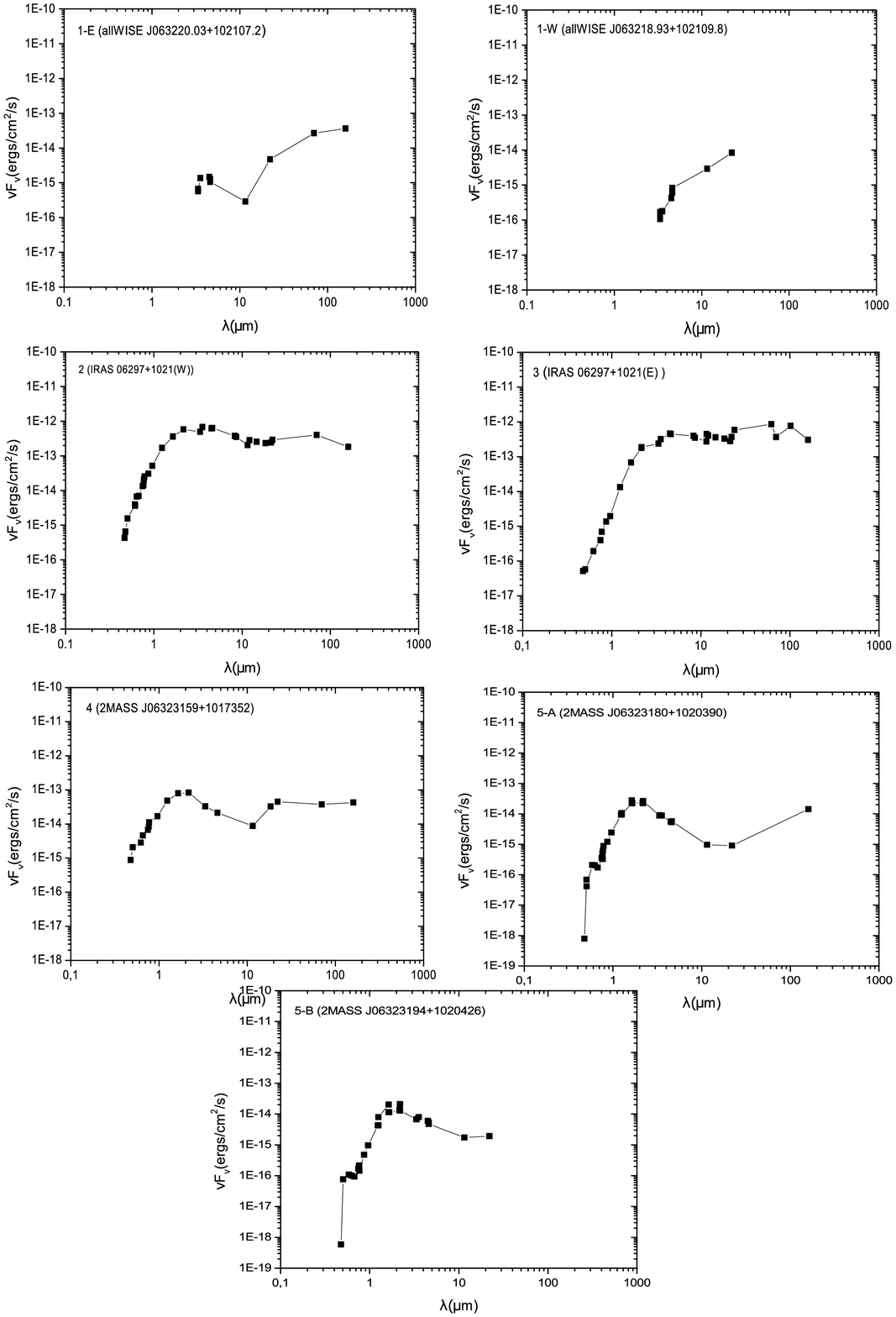}
        \caption{SEDs for the far-IR sources (Table 2, Nos. 6-11).}
        \label{SEDs_1}
\end{figure*}

\begin{figure*}
        \includegraphics[width=400pt]{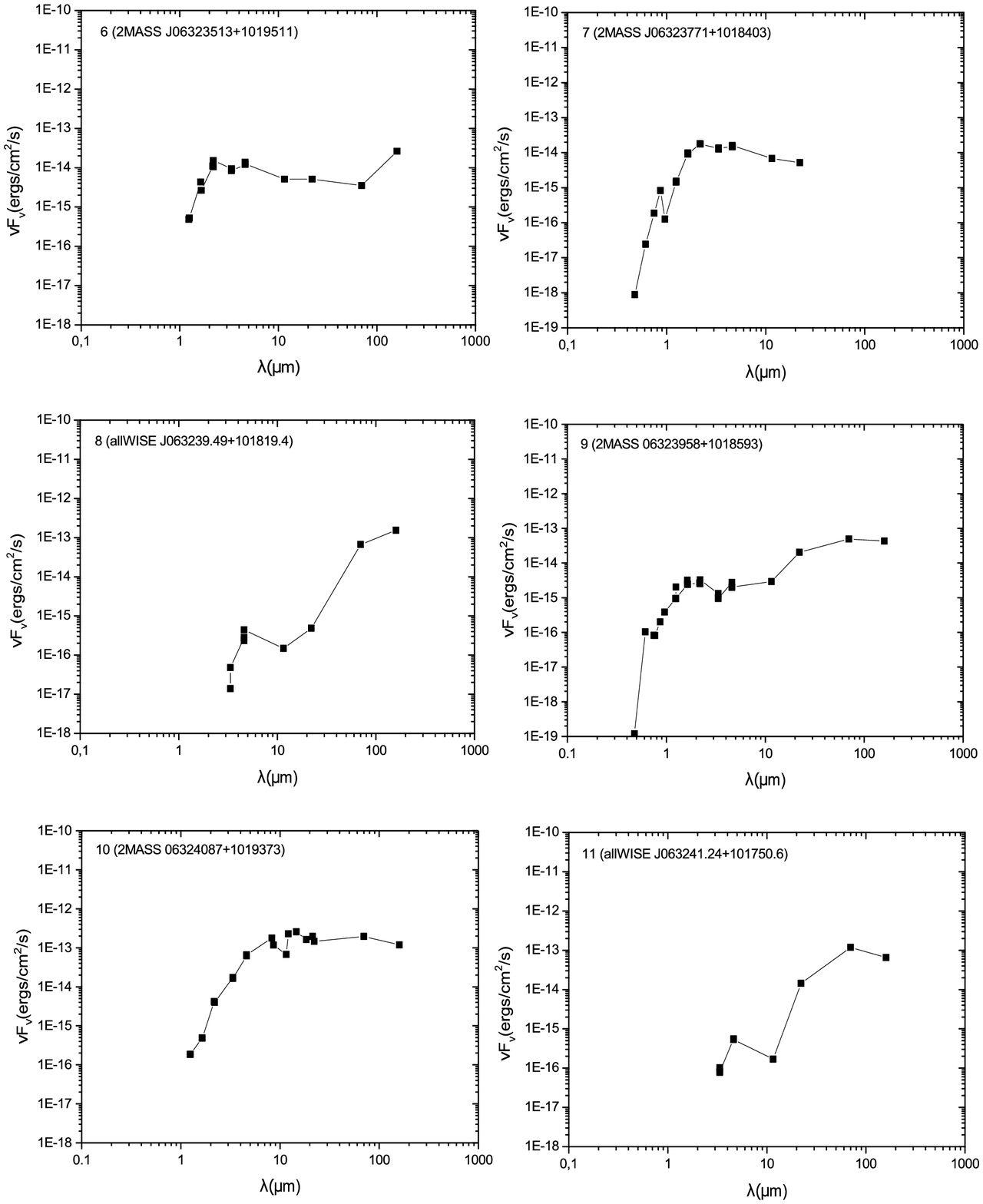}
        \caption{SEDs for the far-IR sources (Table 2, Nos. 6-11).}
        \label{SEDs_2}
\end{figure*}

Therefore we prepared the list of eleven sources, visible in the PACS image (some of them are double), and attempted to identify these far-IR sources   with known objects and with the newly discovered H$_{2}$ flows (which also are shown schematically on Fig.\ref{PACS}).  Using the Vizier photometry viewer and, when available, the data from the PACS Point Source Catalog, we also built spectral energy distributions (SEDs) for the all  sources, found by us.  The results are presented in Table 2, Figs. \ref{SEDs_1} and \ref{SEDs_2} and are discussed below.
Coordinates were measured from the WISE images, excluding the cases when objects were not visible even in the WISE survey.

\textbf{1-W and 1-E}. This is a double source, hidden inside the biconical reflection nebula, probably connected with MHO 3144 (see above). Both sources are invisible in 2MASS. It seems that each of them illuminates
its own lobe of the nebula. The western source is redder and more extended in mid-IR range, but the eastern one is much brighter in far-IR and coincides with PACS PSC coordinates (nevertheless, the western source is faintly visible on PACS images). Thus, we added the data of PACS photometry to the SED of the eastern source. Each star in this pair can be considered as the probable source of MHO 3144
outflow.

\textbf{2 and 3.} These two stars, being  IRAS sources, naturally are brightest in the PACS image. As was already mentioned above,  they are described in detail in our previous paper.
Here we added the PACS photometry to their SEDs.

\textbf{4.} This object is identified with the star 2MASS J06323159+1017352 in the small cometary nebula Petr 7, which, with the previous two objects, was considered as the possible source for the HH 1203 flow \citep{MMD}. The nebula, illuminated by this star, is visible in mid-IR.  The existence of far-IR excess makes this object more interesting than was thought before.

\textbf{5.} A pair of stars, very faint and red in optical range, but prominent in IR.
It seems that far-IR measurements correspond to the southern star, which is brighter in all ranges.
   
\textbf{6.} This star is not visible in optical range, but it is pretty bright in all IR surveys. Our images show that it is surrounded with a reflection comet-like nebula, divided by a narrow streak of an absorbing material. The nebula can be seen also on 2MASS image. This object was (most probably wrongly) considered as a galaxy \citep{glade}.

\textbf{7.} This faint in optical range, but rather bright in near and middle IR star is located just near the southern end of MHO 3143 flow. Thus, it can be considered as its probable source, though there are more attractive candidates (see below). It is faint in far-IR range and has no PACS photometry. 

\textbf{8.} This unusual source is one of the brightest in the PACS image of the area. Nevertheless, no stellar-like object is visible neither in optical, nor in near-IR ranges. On the 4.6 $\mu$m images of allWISE and unWISE surveys as well as in GLIMPSE360 color image, an extended knot can be seen in this location.
This knot  seems to be a part of MHO 3145 flow. It, though very faintly, is present on the our  2.1 $\mu$m image. One can suppose that here we observe the source of the flow, deeply embedded into dust.

\textbf{9.} This source coincides with pair (7.5\arcsec) of relatively bright in near-IR stars; the southern one is visible even in the optical range. These
stars are located on the axis of MHO 3143 flow, and one cannot exclude that one of them can be a possible  source of this flow, though this seems unlikely in view of more probable candidates. In the PACS image the object is elongated;\ thus, both stars must have significant luminosity in far-IR range.

\textbf{10.}
 This source corresponds to the bright and rather red star, well visible in all IR surveys. It is located near the northern  end of   MHO 3143 and seems to be obviously connected with this flow (which, as was mentioned above, abruptly changes direction near this source). Its significant  bolometric luminosity (see below) reinforces this opinion. However, one must bear in mind the presence of other suitable objects, discussed here.  

\textbf{11.} The reddest and probably most unusual source in this area. Actually, besides of PACS survey, it is visible only in allWISE W4 band (22  $\mu$m), and, which is especially interesting, in allWISE and unWISE W2 band (4.6 $\mu$m). This can be an evidence of the presence of H$_2$ emission in its surrounding.

From Figs. \ref{SEDs_1} and \ref{SEDs_2} one can see that SEDs of nearly all described above objects are flat from near-IR to far-IR which points to the wide temperature range in the surrounding dust envelopes. The exception is extremely red objects number 1-E, 1-W, 8 and 11, the SEDs of which constantly rise up to far-IR.

We roughly estimated the bolometric luminosities of all these sources, integrating their SEDs and assuming 715 pc for the distance of the complex. These values are listed in the Table 3.
The luminosities of the objects 2, 3 and 4, for which far-IR measurements were added, should be compared with the estimates presented in our previous paper \citep{MMD}. As one can expect, they slightly increased. Then, the source 10 stands out  by its bolometric luminosity. It is the fourth object in this field with L $\geqslant$\ 10L$_{\sun}$  (besides of three more
luminous HAeBe stars). This star is a very probable source of MHO 3143. Among other objects we should note the sources 8  and 11. Their bolometric luminosities are near 3 L$_{\sun}$; on the other hand, they are very red and are not visible in the optical range. Besides, they are  probable sources of the observed or suspected  H$_2$  outflows. All this makes the sources 8, 10 an 11, as well as nebulous 1-E and 1-W a good target for the further IR studies.
 
\begin{table}
 \centering
 \begin{minipage}{80mm}
  \caption{Luminosities of far-IR sources.}
  \begin{tabular}{ll@{}}
  \hline
   No.     &  L (L$_{\sun}$) \\
 \hline
 1-W & 1 \\
 1-E & 0.1 \\
 2 & 31 \\
 3 &  36 \\
 4 &  3\\
 5a & 1.1   \\
 5b & 0.2 \\
 6 & 0.7 \\
 7 & 0.5\\
 8 & 3\\
 9 & 1.8\\
 10 & 10.4\\
 11 & 3.4  \\ 
 \hline

\end{tabular}
\end{minipage}
\label{lum}
\end{table}  

\section{Conclusion}
 
Recent studies \citep{MMD,bhadari} found that the star formation in Mon R1 association is significantly more active than appeared before and is continuing in the present time. The field, described in the present paper, is a region with enhanced sub-mm emission on 350 and 550 $\mu$m \citep{Montillaud,bhadari}, but it did not became the target of further studies in these papers. Nevertheless, the discovery of a compact cluster of far-IR sources and a group of related to them new H$_2$  outflows inside this area once more indicates significant activity of the  star formation in Mon R1. Comparison of the shapes of SEDS of these IR sources with the samples, presented by \citet{greene}, confirms that this group contains very young PMS stars, probably Class II and Class I sources. Thus, the rate of the current low and intermediate mass star formation in Mon R1 becomes even higher.

\section*{Acknowledgments}

Authors thank referee for very useful suggestions. They want to express their thanks to S.A. Lamzin, whose interest has helped in the organization of these observations.   Also authors thank D. Froebrich for the providing MHO numbers. This work was partly supported by the RA State Committee of Science, in the frames of the research projects 18T-1C-329 and 21T-1C031.
The work of A.Tatarnikov was supported by
the Interdisciplinary Scientific and Educational School of Moscow
University ``Fundamental and Applied Space Research''. This research has made extensive use of Aladin sky atlas, VizieR catalogue access tool and SIMBAD database,
which are developed and operated at CDS, Strasbourg Observatory, France.
PACS has been developed by a consortium of institutes led by MPE (Germany) and including UVIE (Austria); KU Leuven, CSL, IMEC (Belgium); CEA, LAM (France); MPIA (Germany); INAF-IFSI/OAA/OAP/OAT, LENS, SISSA (Italy); IAC (Spain). This development has been supported by the funding agencies BMVIT (Austria), ESA-PRODEX (Belgium), CEA/CNES (France), DLR (Germany), ASI/INAF (Italy), and CICYT/MCYT (Spain). 

\section*{Data availability}

The data underlying this article will be shared on reasonable request to the corresponding author.

\label{lastpage}

\end{document}